\documentclass{ws-rv961x669}
\usepackage{ws-rv-van}     
\usepackage{ws-rv-thm}     
\usepackage{subfigure}     
\makeindex

\index{equations!display}
\begin{document}

\chapter[Jacob Bekenstein's Universes: From Black Holes to
Modified Theories of Gravity]{Jacob Bekenstein's Universes: From Black Holes to
Modified Theories of Gravity\label{ra_ch1}}

\author[Ofer Lahav]{Ofer Lahav}

\address{Department of Physics \& Astronomy\\
University College London, Gower Street, London WC1E 6BT, UK \\
o.lahav@ucl.ac.uk}

\begin{abstract}
Jacob Bekenstein's pioneering contributions to the foundations of Gravity cover a broad range of scales, from 
Black Holes to the entire Universe. He is well known as the father of Black Hole Thermodynamics and the inventor of the TeVeS  modified gravity theory. 
I reflect here on my interaction with him, from the time he supervised me for my MSc project at Ben-Gurion University in the early 1980s to his sabbatical visit at UCL in 2014, about a year before he passed away (sadly just a month before the first LIGO discovery of Gravitational Waves from a Binary Black Hole).
His quote  ``I feel more comfortable in the world if I understand why things are happening"
reflects the way he viewed the world as a Physicist as well as  his personal belief. 
He also allegedly  said ``Sometimes theories are smarter than their inventors". 

\end{abstract}


\body

\bigskip

\section {Introduction} 
On the early morning of 17 August 2015 I was about to catch a Eurostar train to France. A sad email arrived that Jacob Bekenstein had passed away unexpectedly the previous day of a heart attack, while visiting Helsinki. 
He was only 68 years old.  Bekenstein was the father of ``Black Hole Thermodynamics". I had the privilege to be supervised by him for my MSc degree back in Ben-Gurion University in Beer-Sheva Israel in the early 1980s. He visited me at University College London (UCL) on his sabbatical term just a year before he passed away.
Jacob was born in Mexico in 1947. After studying in New-York and Princeton he was a Post-doc in Texas. He made an unusual career move to immigrate to Israel, to the Ben-Gurion University in Beer-Sheba (then the youngest university in the country). 
In recognition of his pioneering work Bekenstain received the Wolf Prize, the Israel Prize, the Einstein Prize and other awards.

\bigskip 

\section {The First Stars in the Universe} 
My MSc thesis with Bekenstein  was not on Black Holes, but on the formation of the first stars in the Universe.
In a memorial service for Jacob at the Israeli Academy of Sciences in Jerusalem (in July 2016) I recalled 
meeting him for the first time.  
After three years studying for a BSc in Physics at Tel-Aviv University, I was in the `Atuda' (a scheme which allows one to study at university first, before doing military 
service).  I served for five years at a base of the Israeli Air force near Beer-Sheba. 
It made good sense, as a way of keeping my sanity in a five year service, to enrol at nearby Ben-Gurion University for an MSc degree in Astrophysics. It was highly recommended to me to have Bekenstein as my project supervisor. On a first meeting with him, in a modest office he shared with David Owen (who later lectured to me),  Bekenstein said he was not available for taking on a student, as he was busy working himself on a research project (later I understood it was about his ideas to modify Einstein's gravity, to overcome the  dark matter problem, see below). I am not sure how I had the courage to approach him again some weeks later, and what made him change his mind. He was actually very generous with his time, and very kind to me.  Taking into account that I was a part-time student, he `cooked' for me a project on a `quiet topic', which he correctly envisaged would become more relevant in years to come. The project was entitled ``cooling processes of formation of the first stars in the Universe".  

\begin{figure}[hb]
\centerline{\includegraphics[width=12cm]{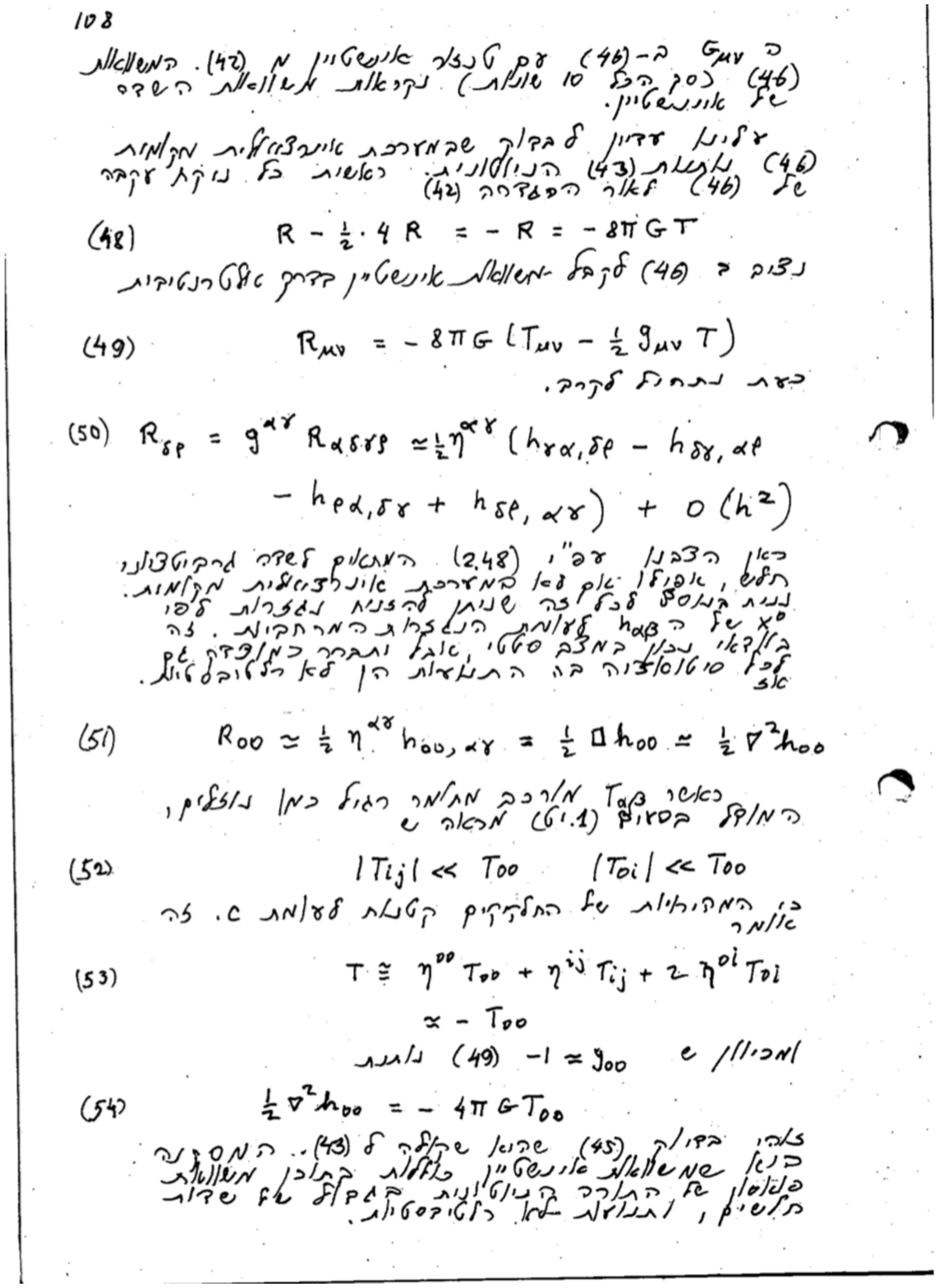}}
 \caption{Notes in Hebrew from Bekenstein's lectures on General Relativity given at the Hebrew University. Taken from his website 
http://www.phys.huji.ac.il/$\sim$bekenste/}
\label{ra_fig1}
\end{figure}

The question was, given the limited chemical reactions possible in the primeval Universe, how could a cloud of gas collapse? The fate of a collapsing cloud depends on the competition between gravity and pressure. For the gravity to win over pressure, the pressure has to be reduced by cooling of the gas. It is believed the cooling can be achieved by molecular hydrogen. 
This work resulted in a paper~\cite{lahav86}~. I was hoping to have Bekenstein as a co-author, but he declined saying he had not done 50\% of the work. Later this became an important topic, with these reactions implemented in sophisticated hydrodynamical simulations. He always emphasized the importance of {\it thinking}  about the science question. 
Straight after completing both the MSc and my service in the Air Force I moved to study for my PhD at the University of Cambridge, where I was mentored by George Efstathiou and by Donald Lynden-Bell (who sadly passed away in February 2018). I then combined Post-doctoral fellowships in Cambridge and in Princeton, before becoming a member of staff  at the Institute of Astronomy in Cambridge. In 2004 I moved to UCL  as Perren Chair of Astronomy, to start there the Cosmology area. Without the foundations I received by working  with Bekenstein I doubt that my career in Astrophysics would have developed in such a  fortunate way.

\section{Black Hole Thermodynamics} 
Bekenstein is recognized as a pioneer of  the idea  of `Black Hole Entropy'.  This fairly abstract concept has led to one of the most important equations in the whole of Physics. This brings together two topics:  The first, Entropy, is of great relevance to daily life. The second, Black Holes, is very remote from human experience. 
Entropy, usually labeled by the letter S, is the measure of `disorder' (or randomness) of a system. 
A common example in explaining Entropy is the way a perfume sprayed in one corner of a room (in an `ordered' state) will eventually be distributed in the entire room (in a `disordered' state). The reverse process has never been observed.
The Second Law of Thermodynamics states that the amount of Entropy (disorder) of an isolated system can never decrease over time.  
Ludwig Boltzmann (1844-1906) formulated Entropy as 
\begin{equation}
 S = k_B \ln (W) \;,
 \label{entropy_S}
 \end{equation} 
 where W is the number of states of the system, and  $k_B$  is Boltzmann's constant.
  This equation appears on Boltzmann's gravestone in Vienna.
 
The other part of the story is that of  Black Holes. 
In short, a Black Hole is a compact region with such strong gravity that nothing can escape from it, not even light. 
What is the connection between Entropy and Black Holes?
Until the early 1970s the accepted picture was that a Black Hole hides everything. John Wheeler, one day in the early 1970s, allegedly asked his then PhD student Jacob Bekenstein, ``What would happen if you poured a cup of hot tea into a Black Hole?".
This question may sound naive, but is actually deep and inspiring.  
Bekenstein rescued the Second Law of Thermodynamics from an embarrassment, by allowing the Entropy of the cup of tea to be converted to the Black Hole Entropy (proportional to the Event Horizon area $A$). This way the Entropy of the Black Hole system never decreases. 

In his influential 1973 paper~\cite{bek73} Bekenstein wrote, ``There are a number of similarities between Black-Hole physics and thermodynamics. Most striking is the similarity in the behaviours of Black-Hole area and of Entropy: both quantities tend to increase irreversibly."
The Black-Hole Entropy was eventually written as:
\begin{equation}
S_{BH} = {{k_{B} A }\over {4 l_{P}^2} } \;,
\label{S_BH}
\end{equation}
where $l_p = {\sqrt {G \hbar /c^3}} $ is the Planck length.
The history of the Bekenstein \& Hawking  iterations  leading to final form is interesting by itself, but beyond the scope of this article. 
I will only add that I recall in the late 1980s a lecture  by Denis Sciama in Cambridge. Sciama (1926-1999),  the father of the modern British cosmology, supervised Stephen Hawking and Martin Rees, among many others.  In that lecture Sciama wrote on the blackboard the equation for $S_{BH}$  and said that the subscripts `BH' may stand for either `Black Hole' or `Bekenstein-Hawking'. The equation for the Black Hole temperature appears on Stephen Hawking's gravestone in Westminster Abbey.
It is not obvious  why the Black Hole Entropy depends on the area (or the boundary) rather that on the volume. This may remind us of Gauss' theorem, that the electric fields integrated over the surface tells us about the total electric charge within the boundary. It is a kind of topological property.  Black Hole Entropy has led physicists to speculate 
that anything that happens in a volume of space-time can be described by the information on the boundary (surface) of that volume. This is called the `Holographic Principle', and Jacob Bekenstein published several papers on this and on the connection to information theory.
 An overview is given in Bekestein's  book\cite{bek03b} and in his Scientific American article~\cite{bek03a} .

\section {The TeVeS Modified Gravity} 

In the early 1980s 
Bekenstein was inspired by Moti Milgrom's idea\cite{milgrom83} to explain flat galaxy rotation curves by modifying Newtonian Dynamics (MOND),  without the need to invoke Dark Matter. 
After various attempts to generalise MOND to a relativistic theory, Bekenstein devised TeVeS (Tensor, Vector, Scalar). It is derived from the action principle, so it obeys conservation laws.  As  a relativistic theory, it can predict effects of gravitational lensing (in both the weak and strong regimes). TeVeS has been praised as remarkable theoretical achievement, extending Einstein's General Relativity.
However, in its simplest form TeVeS has been challenged in explaining for example the Cosmic Microwave Background peaks, 
and the Bullet Cluster.

Sadly, Bekenstein  did not witness the LIGO discovery of Gravitational Waves. The first event GW150914, of a Binary Black Hole, was detected by LIGO\cite{LIGObbh} on 14 September 2015,  just a month after Bekenstein passed away. 
The LIGO \& Virgo detection\cite{LIGObns} of Binary Neutron Star (BNS) GW170817 and electromagnetic followups turned out to be relevant for constraining cosmological modes. Several ground and space telescopes, including our Dark Energy Survey camera\cite{des-gw1, des-gw2}
in Chile captured this Gravitational Wave `after glow'.
In particular, the Fermi Gamma-ray space telescope  detected a flash from BNS system only 1.7 seconds  after the  Gravitational Wave measurement.
Given the distance of about 40 Mpc to the host galaxy NGC4993, this implies that the speed of Gravitational Waves  is close to the speed of light to within one part in $10^{15}$.
In TeVeS the differences between the speed of gravity and light will be due to coupling
of the curvature to other fields\cite{sagi10}. This may put severe constraints on TeVeS in its simplest form, although some other variants of TeVeS may survive that test. 
Indeed, it was shown recently\cite{skordis19}  that  in a previously unknown class of models, inspired by TeVeS,
the speed of gravitational waves equals to the speed of light. More work is needed for a fair comparison of TeVeS and its extensions with current and future observational data\cite{ez17}.


\section {Reflections} 

Bekenstein was very methodological and very precise in both English and in Hebrew, orally as well as in writing.
We told me once how he mastered the Hebrew language, which was not his native language. 
He read the Hebrew version of the Sears\&Zemansky, that was the the standard High-School Physics textbook in my generation.  
As he knew the Physics, he could very easily pick up the Hebrew words... 

It was a pleasure to host Jacob and his wife Bilha in London for four months in 2014, a year before he passed away.
We discussed various science topics, in what seems now as `closing the loop' more than 30 years after working with him back at Ben-Gurion University.

In a popular article\cite{lm14} we quoted different views of philosophers on how paradigm shifts happen.
This included a a reference to the French physicist and philosopher of science 
Duhem (1861-1916), who suggested that scientists follow their `good sense' when facing dilemmas of this kind. Bekenstein commented by email to me on 5th  May 2014:

\bigskip 
{\it Ofer

Thanks.  Interesting paper.  Some examples were not known to me.
Neither some views.  If you ask me, Duhem's is the most sensible point of view.  He was a professional physicist and left a piece of useful physics, the Gibbs-Duhem equation.  Personally I think the other views are stylized and bear
little relation to how scientists actually work and reason.  As for the
central question what to do with Dark Energy  and Dark Matter, as you say the evidence
will eventually be overwhelming one way or another.

Jacob}

\bigskip
This gives an insight to the way Bekenstein considered the way theories describe nature. In an interview in Hebrew 
he said:  ``I feel more comfortable in the world if I understand why things are happening".
This reflects the way he viewed the world as a Physicist and his personal belief. 
Allegedly he also said ``Sometimes theories are smarter than their inventors". 
Jacob Bekenstein is greatly missed in the new era of Gravitational Wave Physics and cosmological surveys of the deep Universe.

\begin{figure}[hb]
\centerline{\includegraphics[width=12cm]{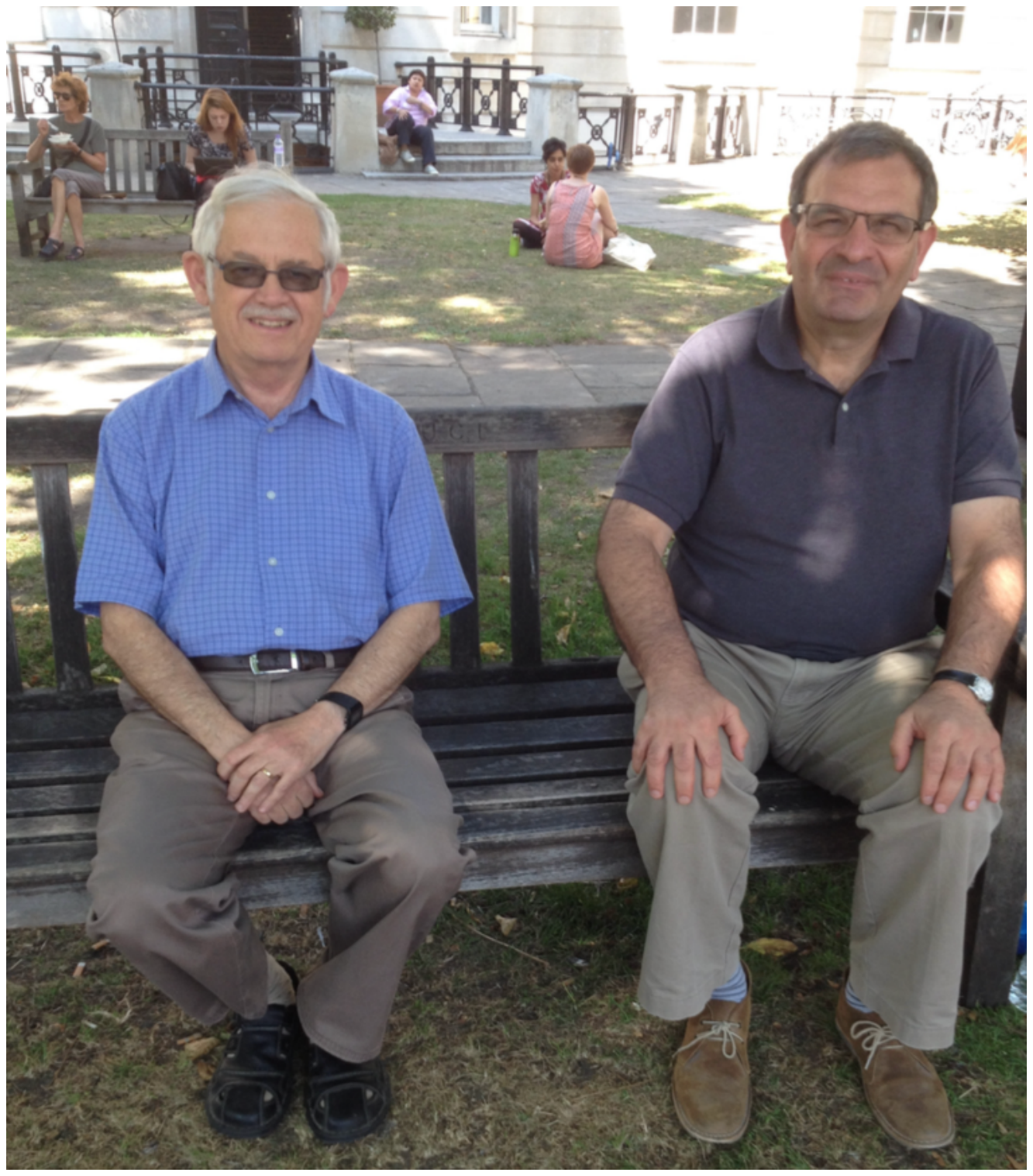}}
\caption{ Jacob Bekenstein and Ofer Lahav  at the main court of UCL,  in July 2014.}
\label{ra_fig2}
\end{figure}

\section{Acknowledgments}
I  thank Ed Copeland, Scott Dodelson, Pedro Ferreira, Moti Milgrom, Jonathan Oppenheim and Tom Zlosnik 
for helpful discussions on the interpretation of Bekenstein's theories in the light of recent observations. 
I also acknowledge support from a European Research Council Advanced Grant FP7/291329 and from the UK Science and Technology Research Council (STFC) Grant No. ST/M001334/1.



\end{document}